\documentclass[aps,prb,preprint,showpacs,preprintnumbers,amsmath,amssymb]{revtex4}
\usepackage{amsmath}
\usepackage{amssymb}
\usepackage{amsfonts}
\usepackage{color,array}
\usepackage{dsfont}
\usepackage{slashed}
\usepackage{graphicx}
\usepackage{graphics}
\usepackage{epsfig}
\usepackage{bbm,bm}
\usepackage{psfrag}
\usepackage{ulem}
\usepackage{url}

\begin{document}

\title{Fisher Exponent from Pseudo-$\epsilon$ Expansion}

\author{A. I. Sokolov}
\email{ais2002@mail.ru}
\author{M. A. Nikitina}
\affiliation{Department of Quantum Mechanics,
Saint Petersburg State University,
Ulyanovskaya 1, Petergof,
Saint Petersburg, 198504
Russia}
\date{\today}

\begin{abstract}
Critical exponent $\eta$ for three-dimensional systems with $n$-vector
order parameter is evaluated in the frame of pseudo-$\epsilon$ expansion
approach. Pseudo-$\epsilon$ expansion ($\tau$-series) for $\eta$ found up
to $\tau^7$ term for $n$ = 0, 1, 2, 3 and within $\tau^6$ order for general
$n$ is shown to have a structure rather favorable for getting numerical
estimates. Use of Pad\'e approximants and direct summation of
$\tau$-series result in iteration procedures rapidly converging to the
asymptotic values that are very close to most reliable numerical estimates
of $\eta$ known today. The origin of this fortune is discussed and shown
to lie in general properties of the pseudo-$\epsilon$ expansion machinery
interfering with some peculiarities of the renormalization group expansion
of $\eta$.

\end{abstract}


\maketitle

\section{Introduction}

Field-theoretical renormalization group (RG) approach proved to be highly
efficient when used to evaluate universal parameters characterizing the
behavior of various systems near Curie point. It yields high-precision
numerical estimates for critical exponents, renormalized coupling
constants, universal ratios, etc., provided lengthy enough RG expansions
are employed and proper resummation of these diverging series is made
(see,. e. g.
Refs.\cite{BNM78,LGZ80,Klei91,AS95,KSF95,GZ97,GZJ98,SOUK99,PS2000,CPV2000}).
Numerical estimates for critical exponents of the Ising, XY, Heisenberg
and some other models obtained within field-theoretical RG machinery are
referred today as canonical numbers\cite{ZJ01,KSF,ZJ,PV02} and widely used
in course of comparison of the theory with computer and physical
experiments including advanced measurements performed in
space\cite{L2003}.

RG expansions being power series in renormalized quartic coupling constant
$g$ or in $\epsilon = 4 - D$ have coefficients that grow factorially with
their number $k$. To struggle this divergency the Borel transformation is
usually employed which turns divergent series into expansions having
non-zero radius of convergence. Resummation methods based on Borel
transformation work very efficiently when original series are alternating
and their coefficients demonstrate regular behavior, i. e. being
monotonically decreasing functions of $k$ for moderate $k$, they
monotonically grow up under $k \to \infty$. Fortunately, RG expansions for
the $\beta$-function and "big" critical exponents ($\gamma$, $\nu$ and
some others) are precisely such regular series both in three and
two\cite{BNM78,LGZ80,OS2000,COPS04} dimensions. This is one of the main
reasons why field-theoretical RG approach turned out to be so effective
numerically in the phase transition problem.

However, there is a critical exponent for which RG series are not so
"friendly". We mean the Fisher exponent $\eta$. Let us look at the
perturbative expansion of $\eta$ for the three-dimensional Ising model
and at the corresponding $\epsilon$-expansion that are known today in
seven-loop\cite{BNM78,MN91} and five-loop\cite{Klei91} approximations
respectively. They are as follows:

\begin{eqnarray}
\eta &=& 0.0109739369 g^2 + 0.0009142223 g^3 + 0.0017962229 g^4
\nonumber\\
&-& 0.000653698 g^5 + 0.00138781 g^6 - 0.0016977 g^7.
\end{eqnarray}
\begin{eqnarray}
\eta &=& 0.01852 \epsilon^2 + 0.01869 \epsilon^3
- 0.00833 \epsilon^4 + 0.02566 \epsilon^5.
\end{eqnarray}

The coefficients of above series as seen to be quite irregular, both in
sign and modulo. That is why the resummation of such series by canonical
(Pade-Borel, conform-Borel, etc.) methods is much less effective than in
the case of Wilson fixed point location $g^*$ and big critical exponents.
As a results, one usually prefers to evaluate the exponent $\eta$ via
scaling relations instead of dealing with corresponding RG series (see,
e. g. comprehensive review\cite{PV02}).

In such a situation it is reasonable to address some alternative technique
which is able to turn original RG expansions into more appropriate ones.
Here we do not mean avoiding of factorial growth of the coefficients since
in series (1), (2) they are small or, at least, not too large. Instead, we
are looking for a tool which would convert RG expansion for $\eta$ into the
series regular in sign along with making higher-order coefficients to
monotonically decrease with growing $k$.

Below, it will be shown that the pseudo-$\epsilon$ expansion can play a
role of such a tool. This approach invented by B. Nickel many years
ago (see Ref. 19 in the paper of Le Guillou and Zinn-Justin \cite{LGZ80})
exploits the idea that Wilson fixed point location in three dimensions may
be found iteratively by means of introducing fictitious small parameter
$\tau$ into linear term of perturbative series for $\beta$-function.
Pseudo-$\epsilon$ expansion proved to be very efficient when used to
estimate critical exponents and other universal quantities characterizing
critical behavior of three-dimensional systems
\cite{LGZ80,GZJ98,FHY2000,HID04,NS14}. Even in two dimensions, where RG
series are shorter and more strongly divergent, it leads to good or
satisfactory numerical results \cite{LGZ80,COPS04,CP05,S2005,NS13}. As we will
see, for the exponent $\eta$ the pseudo-$\epsilon$ expansion turns out to
be highly effective as well: it generates iteration procedures rapidly
converging to the asymptotic values that are in good agreement with
the numbers extracted from alternative field-theoretical and lattice
calculations.

It is worthy to note that ability of pseudo-$\epsilon$ expansion approach
to accelerate RG iterations and to smooth oscillations of numerical
estimates as functions of $k$ was discovered just after beginning of its
application\cite{LGZ80}. It was observed also that in many cases
pseudo-$\epsilon$ expansions do not require advanced resummation
procedures; as a rule, use of Pad\'e approximants or even direct summation
are sufficient to lead to proper numerical results. In our case, however,
pseudo-$\epsilon$ expansion demonstrates an extra advantage -- it turns
the series with a structure rather unfavorable from the computational
point of view into those quite suitable for numerical estimates.

\section{Pseudo-$\epsilon$ expansions for $n = 0, 1, 2, 3$}

Critical thermodynamics of three-dimensional systems with $O(n)$-symmetric
vector order parameters is described by Euclidean field theory with the
Hamiltonian:
\begin{equation}
\label{eq:1}
H = \int d^{3}x \Biggl[{1 \over 2}( m_0^2 \varphi_{\alpha}^2
 + (\nabla \varphi_{\alpha})^2)
+ {\lambda \over 24} (\varphi_{\alpha}^2)^2 \Biggr] ,
\end{equation}
where bare mass squared $m_0^2$ is proportional to the deviation from mean
field transition temperature. Perturbative expansions for the
$\beta$-function and critical exponents of this model were calculated in
the six-loop approximation within the massive theory\cite{BNM78,Guelph}.
Later, RG series for critical exponents were extended up to seven-loop
order by Murray and Nickel in their unpublished work\cite{MN91};
seven-loop terms were reported in the paper of Guida and
Zinn-Justin\cite{GZJ98}.

We derive the pseudo-$\epsilon$ expansion for critical exponent $\eta$
starting from RG series mentioned. To do this one has to substitute the
$\tau$-series for the Wilson fixed point location $g^*$ into perturbative
expansion for the Fisher exponent and reexpand it in $\tau$.
Pseudo-$\epsilon$ expansion of $g^*$ for general $n$ is known up to
$\tau^6$ term (six-loop order)\cite{NS14}. At first glance, with this
expansion in hand $\tau$-series for critical exponents may be found within
the same -- $\tau^6$ -- approximation. It is really so for all critical
exponents but the Fisher one. Since the first non-zero term in RG
expansion of $\eta$ is proportional to $g^2$ the length of $\tau$-series
for $g^*$ turns out to be sufficient to find $\tau^7$ term. Seven-loop
contribution in RG expansion of $\eta$ was calculated for concrete values
of $n$\ most interesting from the physical point of view\cite{MN91}. That
is why here we present pseudo-$\epsilon$ expansions of $\eta$ for $n = 0,
1, 2, 3$ only, leaving six-loop ($\tau^6$) series at generic $n$ for
Section IV. Seven-loop $\tau$-series obtained are as follows:

\begin{eqnarray}
\eta &=& 0.0092592593 \tau^2 + 0.0089160938 \tau^3 + 0.004342287 \tau^4
\nonumber\\
&+& 0.002834158 \tau^5 + 0.00093920 \tau^6 + 0.0017464 \tau^7, \qquad  n = 0
\end{eqnarray}
\begin{eqnarray}
\eta &=& 0.0109739369 \tau^2 + 0.0101871237 \tau^3 + 0.005044182 \tau^4
\nonumber\\
&+& 0.003205816 \tau^5 + 0.00129547 \tau^6 + 0.0017339 \tau^7, \qquad  n = 1
\end{eqnarray}
\begin{eqnarray}
\eta &=& 0.0118518519 \tau^2 + 0.0105390747 \tau^3 + 0.005188190 \tau^4
\nonumber\\
&+& 0.003229563 \tau^5 + 0.00145159 \tau^6 + 0.0016264 \tau^7, \qquad  n = 2
\end{eqnarray}
\begin{eqnarray}
\eta &=& 0.0122436486 \tau^2 + 0.0104041740 \tau^3 + 0.005026652 \tau^4
\nonumber\\
&+& 0.003060806 \tau^5 + 0.00145632 \tau^6 + 0.0014657 \tau^7, \qquad  n = 3
\end{eqnarray}

Series (4)-(7) are seen to have much more regular structure than original RG
expansions. Their coefficients possess the same sign and monotonically
decrease with increasing $k$, apart from those of seven-loop ($\tau^7$)
terms. These coefficients being small are nevertheless some bigger than
their six-loop ($\tau^6$) counterparts signalizing that $\tau$-series remain
divergent. Despite of this, expansions (4)-(7) turn out to be quite suitable
for getting numerical estimates.

\section{Numerical results: fast convergence to accurate asymptotes.}

Numerical values of $\eta$ are extracted from the series (4)-(7) by means
of Pad\'e approximants [L/M] and by direct summation (DS). Pad\'e
triangles for Ising and Heisenberg models are presented in Tables I and II
as typical examples. Note that symbol [L/M] denotes here Pad\'e
approximants constructed for $\eta/\tau^2$, i. e. with insignificant
factor $\tau^2$ having physical value $\tau = 1$ ignored.

As seen from Tables I and II the estimates of $\eta$ given by
highest-order near diagonal approximants [3/2] and [2/3] are very close to
the numbers resulting from resummed 3D RG expansions\cite{GZJ98,JK01};
they differ from each other by 0.001 (3 per cent) or less. Moreover, the
convergence of pseudo-$\epsilon$ expansion estimates to the asymptotic
values turns out to be fast what also may be seen from both tables.
Similar behavior of estimates is observed in the case of direct summation
of the series (4)-(7).

This is clearly demonstrated by Table III where Pad\'e and DS
estimates of $\eta$ as functions of $k$ are collected along with
the values the field-theoretical and lattice calculations yield.
One can see that for all four values of $n$ both iteration schemes
lead to the numbers which agree well with other high-precision
estimates. As seen from Table III the deviation of
pseudo-$\epsilon$ expansion estimates from the alternative values
is much smaller than characteristic difference between these
values themselves. DS estimates reach their asymptotes
monotonically what is a direct consequence of the $\tau$-series
structure. On the contrary, the behavior of Pad\'e estimates turns
out to be oscillatory, i. e. typical for this and other, more
sophicticated resummation procedures (see, e. g. classical papers
\cite{BNM78,LGZ80}). Corresponding oscillation, however, are weak
what is known to be specific for the pseudo-$\epsilon$ expansion
technique.

Keeping in mind optimistic results just obtained, the question arises: will
numerically favorable structure of pseudo-$\epsilon$ expansion for Fisher
exponent demonstrated at $0 \le n \le 3$ persist for larger $n$? In other
words, whether numerical power of the pseudo-$\epsilon$ expansion is its
generic property or it manifests itself only for moderate $n$? To answer
this questions we are in a position to study $\tau$-series for $\eta$ at
arbitrary $n$.

\section{Large $n$ and roots of fortune.}

Perturbative RG expansion of $\eta$ for general $n$ are known today within
six-loop approximation\cite{AS95}. This enables us to derive corresponding
pseudo-$\epsilon$ expansion ranging up to $\tau^6$ term. Straightforward
calculation leads to the following $\tau$-series:
\begin{eqnarray}
\label{eta-tau}
\eta&=& {\tau^2 \over (n + 8)^2} \biggl(0.5925925926 + 0.2962962963 n \biggr)
\nonumber \\
&+& {\tau^3 \over (n + 8)^4} \biggl(36.52032036 + 26.24895084 n + 4.043763332 n^2
+ 0.0246840014 n^3 \biggr)
\nonumber \\
&+& {\tau^4 \over (n + 8)^6} \biggl(1138.304360 + 1139.876143 n + 362.9490746 n^2
\nonumber \\
&+& 39.31000643 n^3 + 0.2496327902 n^4 - 0.0042985626 n^5 \biggr)
\nonumber \\
&+& {\tau^5 \over (n + 8)^8} \biggl(47549.2884 + 57808.8268 n + 26407.6964 n^2
+ 5708.39224 n^3
\nonumber \\
&+& 519.915765 n^4 + 6.06899481 n^5 - 0.3213367385 n^6
- 0.006550922 n^7 \biggr)
\nonumber \\
&+& {\tau^6 \over (n + 8)^{10}} \biggl(1008457.5 + 1750566.0 n + 1226035.8 n^2
+ 440973.33 n^3
\nonumber \\
&+& 83719.223 n^4 + 7199.2401 n^5 + 92.760879 n^6
\nonumber \\
&-& 10.569441 n^7 - 0.41561284 n^8 - 0.00554892 n^9 \biggr). \ \
\end{eqnarray}
Analyzing this series under various $n$ lying between 4 and 64 we find that:

i) series (8) have positive and monotonically decreasing coefficients up to $n = 24$;

ii) coefficients of $\tau^5$ and $\tau^6$ terms change their signs at $n = 40$ and
$n = 24$ respectively while other coefficients remain positive and monotonically
decreasing;

iii) up to $n = 64$ coefficients of $\tau^5$ and $\tau^6$ terms persist to be
tiny ($\approx 0.0007$ and smaller), so these terms do not influence appreciably
upon numerical estimates the series (8) yields.

Hence, the structure favorable for getting numerical estimates is a generic property
of the pseudo-$\epsilon$ expansion for Fisher exponent. Comparison of the values of
$\eta$ resulting from expansions (4)-(8) with each other and with their counterparts
obtained within other approaches confirms this conclusion. These values are collected
in Table IV, along with the numbers given by the $1/n$-expansion\cite{VPH82}:
\begin{equation}
\eta = {8 \over {3 \pi^2}}{1 \over n} - {512 \over {27 \pi^4}} {1 \over n^2}
- {1.881234507 \over n^3} \ \ .
\end{equation}
All the data presented are seen to be in a good agreement at any $n$.

Why the pseudo-$\epsilon$ expansion technique turns out to be so efficient
in particular case considered? We have an explanation of this fact.
The point is that mechanism of pseudo-$\epsilon$
expansion is organized in such a way that it suppresses the divergency of RG
expansions for critical exponents provided these expansions are alternating.
The mechanism of suppression
works well for alternate series because in course of transformation of RG series
into pseudo-$\epsilon$ expansions multiple mutual subtractions ("destructive
interference") of the terms of original series take place. If, however, we apply
this technique to series with positive coefficients the subtraction is changed
by summation ("contructive interference") what makes relevant terms in
pseudo-$\epsilon$ expansion larger than that of RG series.

This can be demonstrated explicitly. Let the pseudo-$\epsilon$ expansion for
renormalized quartic coupling constant at criticality (Wilson fixed point location)
$g^*$ be:
\begin{equation}
g^* = \tau + A \tau^2 + B \tau^3 + C \tau^4 + D \tau^5 + ... ,
\end{equation}
while RG series for some critical exponent $\psi$ have a form:
\begin{equation}
\psi = p_0 - p_1 g + p_2 g^2 - p_3 g^3 + p_4 g^4 - p_5 g^5 + ... .
\end{equation}
Typically, all $p_i$ are positive, i. e. the series (11) is alternating. To obtain
$\tau$-series for $\psi$, we have to substitute expansion (10) into (11). It yields:
\begin{equation}
\psi = p_0 - p_1 \tau + (-A p_1 + p_2) \tau^2 + (-B p_1 + 2A p_2 - p_3) \tau^3
+ [-C p_1 + (A + 2B) p_2 - 3A p_3 + p_4)] \tau^4 + ...
\end{equation}
If coefficients of pseudo-$\epsilon$ expansion (10) for $g^*$ are positive, what is
really the case for several lower-order terms\cite{S2005,NS14}, coefficients of
the series (11) interfere within (12) destructively. It is clearly seen from the
structure of series (12). The character of interference, however, depends crucially
on signs of coefficients of initial RG expansion. Indeed, if we changed signs of
odd terms in (11), i. e. made all terms in (11) positive, destructive interference
would turn into constructive what again is cleary seen from (12).

What happens in the case of Fisher exponent? Since lower-order terms
in RG expansions for $\eta$ have the same sign (see, e. g. (1)) pseudo-$\epsilon$
expansion machine grows them up. On the contrary, for the higher-order terms this
machine works as suppressive because starting from $g^4$ term series (1) becomes
(looks as) alternating. As a result, pseudo-$\epsilon$ expansion technique
transforms RG series with small and irregular coefficients into $\tau$-series
which possesses larger lower-order coefficients and decreasing with $k$
higher-order ones, i. e. demonstrates behavior similar to that of converging
series. Fig.1 illustrates such metamorphosis for the Ising ($n = 1$) and Heisenberg
($n = 3$) models.

In fact, pseudo-$\epsilon$ expansion does not generate convergent expansions.
Instead, it replaces one diverging series by another, less strongly divergent.
The resulting expansions, however, have much more favorable structure from
the numerical point of view. In our case pseudo-$\epsilon$ expansions actually do
not require resummation, even in its simplest -- Pad\'e -- form: as seen from
Tables III and IV the highest-order ($\tau^7$) estimates given by direct
summation and found by means of Pad\'e analysis differ from each other by 3
percents or less. This difference is much smaller than individual and overall
error bars characteristic for data provided by high-precision field-theoretical
and lattice calculations. Thus, as was argued earlier\cite{NS14,NS13}, the
pseudo-$\epsilon$ expansion approach may be considered as a resummation method.
This method, however, is somewhat specific -- it does not turn divergent series
into convergent but makes them very convenient for practical use.

\section{Conclusion}

To summarize, we have calculated pseudo-$\epsilon$ expansions of the Fisher
critical exponent up to $\tau^7$ terms for $n = 0, 1, 2, 3$ and within six-loop
($\tau^6$) approximation for general $n$. These expansions have been found to
possess a structure that is rather favorable for getting numerical estimates.
Having processed $\tau$-series obtained by means of Pad\'e approximants and
performed their direct summation we've obtained numerical estimates of $\eta$
that are as accurate as those extracted from advanced field-theoretical and
lattice calculations. The structure of $\tau$-series for $\eta$ persists to be
favorable within the wide range of $n$ signaling that it is a generic property
of the pseudo-$\epsilon$ expansion for $\eta$. We have found arguments shedding
light on the roots of such fortune. They lie in the general properties of the
pseudo-$\epsilon$ expansion machinery interfering with some peculiarities of the
RG expansion for $\eta$.

\newpage

\begin{table}[t]
\caption{Pad\'e triangle for pseudo-$\epsilon$ expansion of critical
exponent $\eta$ of 3D Ising model. Approximants are constructed for
$\eta/\tau^2$, i. e. with factor $\eta/\tau^2$ omitted. Approximants
[0/1], [0/3], [0/5] and [4/1] have poles close to 1; their locations
are shown as subscripts. Canonical values\cite{GZJ98} of $\eta$
resulting from resummed RG series and $\epsilon$-expansion are
$0.0335 \pm 0.0025$ and $0.0365 \pm 0.0050$ respectively.}
\label{tab1}
\renewcommand{\tabcolsep}{0.4cm}
\begin{tabular}{{c}|*{6}{c}}
$M \setminus L$ & 0  & 1 & 2 & 3 & 4 & 5  \\
\hline
0 & 0.0110           & 0.0212 & 0.0262 & 0.0294 & 0.0307 & 0.0324  \\
1 & 0.1531$_{1.08}$ & 0.0312 & 0.0350 & 0.0316 & 0.0256$_{0.75}$ &         \\
2 & 0.0232           & 0.0339 & 0.0327 & 0.0333 &        &         \\
3 & 0.0467$_{1.25}$ & 0.0322 & 0.0332 &        &        &         \\
4 & 0.0258           & 0.0342 &        &        &        &         \\
5 & 0.0570$_{1.13}$ &        &        &        &        &         \\
\end{tabular}
\end{table}

\begin{table}[t]
\caption{The same as Table I but for $n = 3$ (Heisenberg model).
Approximant [4/1] has a pole practically equal to 1; corresponding
estimate does not exist. Resummed 3D RG series give $0.0355 \pm 0.0025$
\cite{GZJ98} and $0.0350 \pm 0.0008$\cite{JK01} while the
$\epsilon$-expansion yields $\eta$ = 0.0355\cite{GZJ98}.}
\label{tab2}
\renewcommand{\tabcolsep}{0.4cm}
\begin{tabular}{{c}|*{6}{c}}
$M \setminus L$ & 0 & 1 & 2 & 3 & 4 & 5 \\
\hline
0 & 0.0122 & 0.0226 & 0.0277 & 0.0307 & 0.0322 & 0.0337 \\
1 & 0.0815$_{1.18}$ & 0.0324 & 0.0355 & 0.0335 & - &     \\
2 & 0.0265 & 0.0346 & 0.0341 & 0.0347 & & \\
3 & 0.0414 & 0.0339 & 0.0344 & & & \\
4 & 0.0304 & 0.0354 & & & & \\
5 & 0.0432$_{1.25}$ & & & & & \\
\end{tabular}
\end{table}

\begin{table}[t]
\caption{Convergence of two iteration schemes generated by the
pseudo-$\epsilon$ expansion of critical exponent $\eta$ for the polymer
(SAW), Ising, XY and Heisenberg models; $k$ is the order of approximation
(number of loops). Upper lines contain estimates obtained with a help of
Pad\'e approximants, lower lines correspond to direct summation. Pad\'e
estimates are those given by diagonal approximants [1/1], [2/2] for
$\eta/\tau^2$ or by near diagonal ones; in the latter case Pade estimates
being the averages over two working approximants. Since approximant [0/1]
has pole close to 1 its counterpart [1/0] is used for final estimates;
they are marked with asterisks. High-precision values of Fisher exponent
resulting from 6-loop RG series, 5-loop $\epsilon$-expansion and lattice
calculations (LC) are presented for comparison.} \label{tab3}
\begin{tabular}{c|cccccc|ccc}
\hline
~~$k$~~ & 2 & ~~3~~ & ~4~ & ~5~ & 6 & 7 & 3D RG & $\epsilon$-exp.& LC \\
\hline
\multicolumn{10}{c}{$n=0$} \\
\hline Pad\'e & ~0.0093~ & 0.0182$^*$ & ~0.0266~ & ~0.0300~ & ~0.0280~ &
~0.0285~ &
~0.0284\cite{GZJ98}~  & ~0.0315\cite{GZJ98}~ & \\
\hline
DS & ~0.0093~ & ~0.0182~ & ~0.0225~ & ~0.0254~ & ~0.0263~ & ~0.0280~ &  &  & \\
\hline
\multicolumn{10}{c}{$n=1$} \\
\hline Pad\'e &~~0.0110~~& 0.0212$^*$
&~~0.0312~~&~~0.0344~~&~~0.0327~~&~~0.0332~~&
~0.0335\cite{GZJ98}~ & ~0.0365\cite{GZJ98}~ & 0.0360\cite{CPRV2002,BC2000,BC2002} \\
\hline DS
&~~0.0110~~&~~0.0212~~&~~0.0262~~&~~0.0294~~&~~0.0307~~&~~0.0324~~
& ~0.0335\cite{JK01} &   & 0.0364\cite{PV02} \\
\hline
\multicolumn{10}{c}{$n=2$} \\
\hline Pad\'e & ~0.0119 & ~0.0224$^*$~ & ~0.0326~ & ~0.0356~ & ~0.0343~ &
~0.0348~ & ~0.0354\cite{GZJ98}~ & ~0.0370\cite{GZJ98}~ & 0.0380\cite{CHPRV2001}\\
\hline DS   & ~0.0119~ & ~0.0224~ & ~0.0276~ & ~0.0308~ & ~0.0323~ &
~0.0339~ &
~0.0349\cite{JK01}~ &      &   \\
\hline
\multicolumn{10}{c}{$n=3$} \\
\hline Pad\'e & ~0.0122~ & 0.0226$^*$ & ~0.0324~ & ~0.0351~ & ~0.0341~ &
~0.0346~ & ~0.0355\cite{GZJ98}~ & ~0.0355\cite{GZJ98}~ & 0.0375\cite{CHPRV2002} \\
\hline DS    & ~0.0122~ & ~0.0226~ & ~0.0277~ & ~0.0307~ & ~0.0322~ &
~0.0337~
& ~0.0350\cite{JK01}~ &  &  \\
\hline
\end{tabular}
\end{table}

\begin{table}[t]
\caption{Numerical values of the Fisher exponent for various $n$
obtained from seven-loop ($n = 0, 1, 2, 3$) and six-loop ($n \ge
4$) pseudo-$\epsilon$ expansions processed with a help of Pad\'e
approximants and by direct summation (DS). Pad\'e estimates are
those given by diagonal approximants [2/2] for $\eta/\tau^2$ or by
near diagonal ones (for $n = 0, 1, 2, 3$); in the latter case Pade
estimates being the averages over working approximants [3/2] and
[2/3]. At $n$ = 64 and 48 approximant [2/2] has pole close to 1;
the values reported (marked with asterisks) are averages over
numbers given by approximants [3/1] and [1/3]. The values of
$\eta$ resulting from 6-loop RG series in three
dimensions\cite{AS95,GZJ98}, obtained within the
$\epsilon$-expansion\cite{GZJ98} and $(1/n)$-expansion approaches
are presented for comparison.} \label{tab4}
\begin{tabular}{*{7}{c}}\hline
~~$n$~~~~~ & ~~Pad\'e~~ & ~~DS~~ & ~~3D RG\cite{AS95}~~ & ~~3D RG\cite{GZJ98}~~
&~~$\epsilon$-exp.\cite{GZJ98}~~ & ~$(1/n)$-exp.~ \\
\hline
0~~~  & ~~0.0285~   & ~~0.0280~  &           & ~~0.0284~~ & ~~0.0315~~ &  \\
1~~~  & ~~0.0332~   & ~~0.0324~  & ~~0.038~~ & ~~0.0335~~ & ~~0.0365~~ &  \\
2~~~  & ~~0.0348~   & ~~0.0339~  & ~~0.039~~ & ~~0.0354~~ & ~~0.0370~~ &  \\
3~~~  & ~~0.0346~   & ~~0.0337~  & ~0.038~   & ~~0.0355~~ & ~~0.0355~~ &  \\
4~~~  & ~~0.0329~   & ~~0.0313~  & ~0.036~   & ~~0.0350~~ & ~~0.033~~  & ~~0.0260~ \\
8~~~  & ~~0.0260~   & ~~0.0252~  & ~0.027~   &          &           & ~~0.0271~ \\
16~~~ & ~~0.0164~   & ~~0.0163~  & ~0.017~   &          &           & ~~0.0157~ \\
24~~~ & ~~0.01173   & ~~0.01168  & ~0.012~   &          &           & ~~0.01078~ \\
32~~~ & ~~0.00915   & ~~0.00902  & ~0.009~   &          &           & ~~0.00820~ \\
40~~~ & ~~0.00761   & ~~0.00732  &           &          &           & ~~0.00660~ \\
48~~~ & ~~0.00627$^*$  & ~~0.00616  &           &          &        & ~~0.00553~ \\
64~~~ & ~~0.00490$^*$  & ~~0.00466~ &           &          &        & ~~0.00417~ \\
\hline
\end{tabular}
\end{table}

\begin{figure}
\begin{center}
\includegraphics[width=\linewidth]{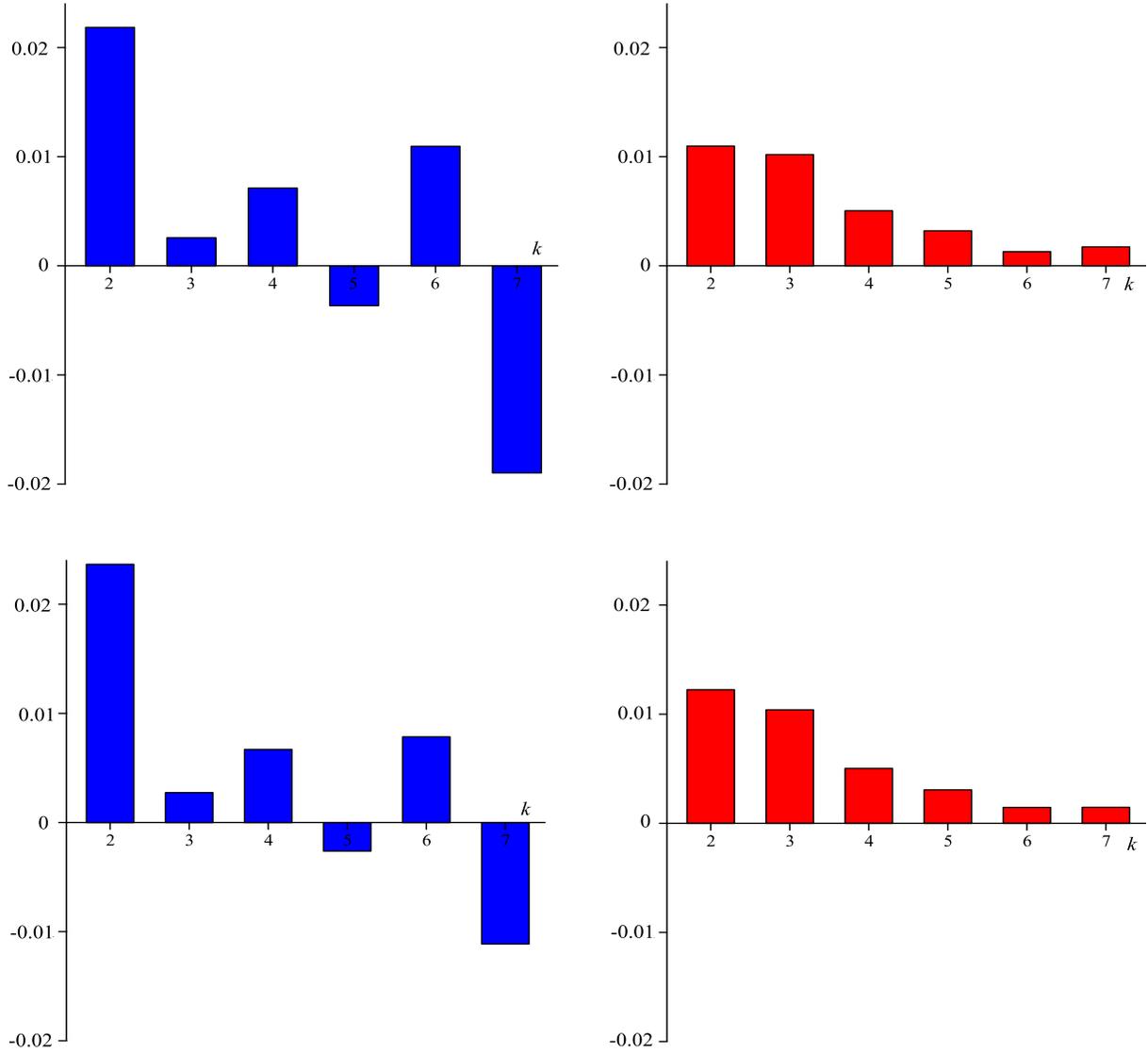}
\caption{(Color online) Fairy metamorphosis of perturbative RG series for
Fisher exponent at $n = 1$ and $n = 3$ under the action of
pseudo-$\epsilon$ expansion machinery. Left hystograms depict weight of
different terms in RG series at the Wilson fixed point, right hystograms
show analogous distributions for pseudo-$\epsilon$ expansions (5) and (7),
$k$ being an order of a term.} \label{fig1}
\end{center}
\end{figure}

\end{document}